\documentclass[sigconf]{acmart}

\usepackage{booktabs} 
\usepackage{multirow}
\usepackage[boxed]{algorithm2e}
\usepackage{verbatim}
\usepackage{setspace}
\usepackage{geometry}
\usepackage{amsmath}
\usepackage{amsfonts}

\begin{document}

\copyrightyear{2018} 
\acmYear{2018} 
\setcopyright{acmcopyright}
\acmConference[SIGIR '18]{The 41st International ACM SIGIR Conference on Research and Development in Information Retrieval}{July 8--12, 2018}{Ann Arbor, MI, USA}
\acmBooktitle{SIGIR '18: The 41st International ACM SIGIR Conference on Research and Development in Information Retrieval, July 8--12, 2018, Ann Arbor, MI, USA}
\acmPrice{15.00}
\acmDOI{10.1145/3209978.3210010}
\acmISBN{978-1-4503-5657-2/18/07}

\title{Explainable Recommendation via Multi-Task Learning in Opinionated Text Data}


\author{Nan Wang, Hongning Wang, Yiling Jia, Yue Yin}
\affiliation{%
  \institution{University of Virginia \\ Department of Computer Science}
  \city{Charlottesville}
  \state{VA}
  \postcode{22904}
}
\email{ {nw6a,hw5x,yj9xs,yy7da}@virginia.edu}

\renewcommand{\shortauthors}{Nan Wang, Hongning Wang, Yiling Jia, Yue Yin}
\renewcommand{\shorttitle}{Explainable Recommendation via Multi-Task Learning}


\begin{abstract}
Explaining automatically generated recommendations allows users to make more informed and accurate decisions about which results to utilize, and therefore improves their satisfaction. In this work, we develop a multi-task learning solution for explainable recommendation. Two companion learning tasks of \textit{user preference modeling for recommendation} and \textit{opinionated content modeling for explanation} are integrated via a joint tensor factorization. As a result, the algorithm predicts not only a user's preference over a list of items, i.e., recommendation, but also how the user would appreciate a particular item at the feature level, i.e., opinionated textual explanation. Extensive experiments on two large collections of Amazon and Yelp reviews confirmed the effectiveness of our solution in both recommendation and explanation tasks, compared with several existing recommendation algorithms. And our extensive user study clearly demonstrates the practical value of the explainable recommendations generated by our algorithm.
\end{abstract}


\maketitle

\section{Introduction}\label{intro}

Extensive research effort has been devoted to improving the effectiveness of recommendation algorithms \cite{Sarwar:2001:ICF:371920.372071,herlocker2004evaluating,Koren:2009:MFT:1608565.1608614,Ma:2008:SSR:1458082.1458205}; but one fundamental question of \textit{``how a system should explain those recommendations to its users''} has not received enough attention \cite{Zhang:2014:EFM:2600428.2609579}. The lack of transparency \cite{sinha2002role} leaves users in a dilemma: a user can only assess the recommendation quality by taking the suggested actions, e.g., purchase the top-ranked items;
however, in order for him/her to adopt the system's customized results, he/she needs to first build trust over the system. Explaining the automatically generated recommendations would bridge the gap. Arguably, the most important contribution of explanations is not to convince users to accept the customized results (i.e., promotion), but to allow them to make more informed and accurate decisions about which results to utilize (i.e.,  satisfaction) \cite{bilgic2005explaining}.

Existing recommendation algorithms emphasize end-to-end optimization of performance metrics, such as Root-Mean-Square Error and Normalized Discounted Cumulative Gain, which are defined on numerical ratings or ranking orders reflecting a user's overall preference over a set of items. Various algorithms such as collaborative filtering \cite{Baltrunas:2011:MFT:2043932.2043988,Sarwar:2001:ICF:371920.372071,5197422,doi:10.1137/1.9781611972764.58,Salakhutdinov:2007:PMF:2981562.2981720} and factorization based methods \cite{5197422,Ma:2008:SSR:1458082.1458205,Rendle:2011:FCR:2009916.2010002} have been proposed to optimize those metrics. However, it is known that humans are complex autonomous systems: a click/purchase decision is usually a composition of multiple factors. 
The end-to-end learning scheme can hardly realize the underlying reasons behind a user's decision making process. As a result, although such algorithms achieve great success in quantitative evaluations, they are still computerized oracles that merely give advice, but cannot be questioned, especially when a new or wrong recommendation has been made. 
This greatly limits the practical value of such recommendation algorithms.



\begin{figure}[!h]
\vspace{-3mm}
\includegraphics[width=.98\linewidth]{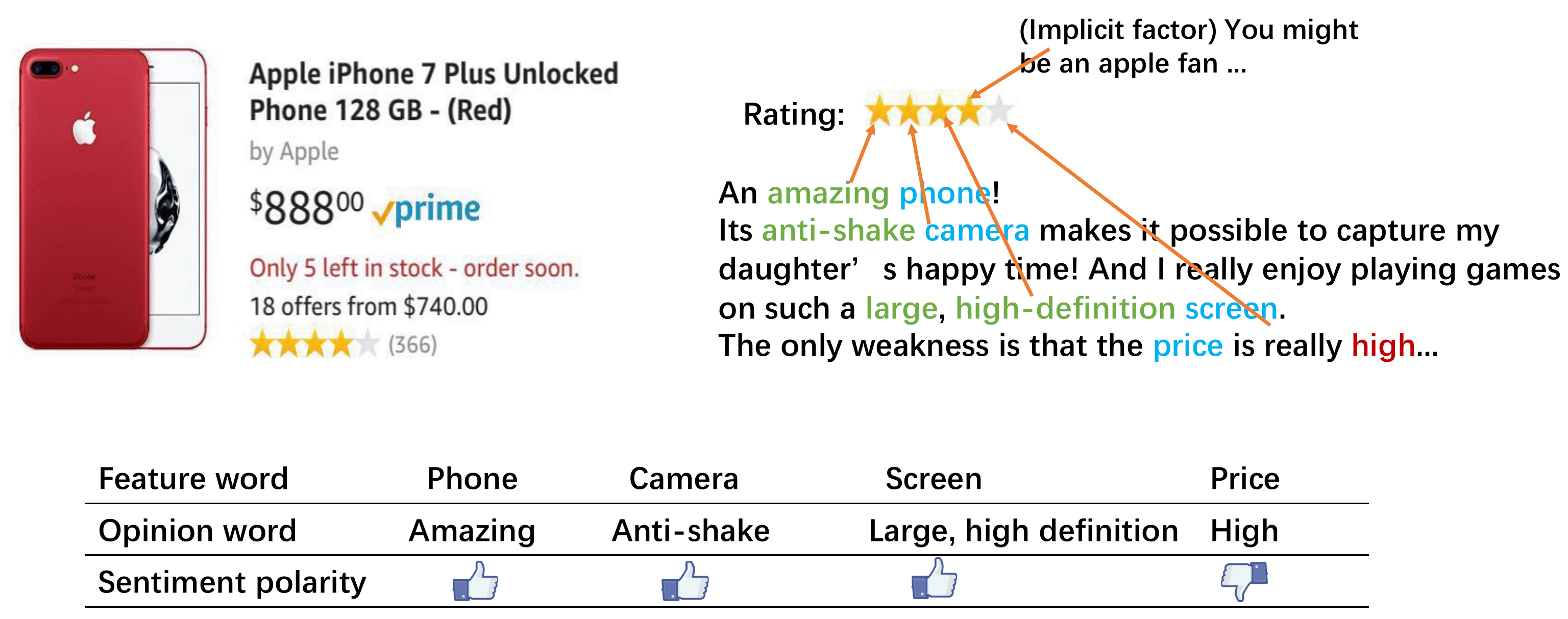}
\vspace{-2mm}
\caption{Composition of star ratings in a typical user review.}
\vspace{-3mm}
\label{fig:compositions}
\end{figure}

We argue that a good recommendation algorithm should consist of companion learning tasks focusing on different aspects of users' decisions over the recommended items, such that the observed final decisions (e.g., clicks or ratings) can be mutually explained by the associated observations. In this work, we focus on opinionated review text that users provide in addition to their overall assessments of the recommended items, and aim to exploit such information to enhance and explain the recommendations. 
Figure \ref{fig:compositions} illustrates how opinionated content in a user's review reflects the composition of his/her overall assessment (rating) of the item. For this given four-star overall rating, three positively commented features contribute to his/her positive assessment, and one negatively commented feature explains his/her negative judgment. If an algorithm could learn to predict not only the user's overall rating, but also the opinionated comment he/she would provide on this item (e.g., how would he/she endorse the features of this item), the recommendation will become self-explanatory. And clearly these two predictions are not independent: the predicted overall assessment has to be supported by the predicted opinionated comments. Therefore, the additional information introduced by the companion content modeling task would help improve the quality of recommendation task.  



Considerable effort has been devoted to utilizing user-generated opinionated content for providing text-based explanations. One typical solution is to combine rating prediction with topic models \cite{McAuley:2013:HFH:2507157.2507163,Xiao2017,Zheng:2016:TTF:3005785.3005923}. 
However, they fail to recognize the detailed opinions on each specific feature or topic, which inevitably leads to biased recommendations or wrong explanations, when a user criticizes a particular feature in an overall positive review. Another type of solutions leverage phrase-level sentiment analysis \cite{Zhang:2014:EFM:2600428.2609579,Ren:2017:SCV:3018661.3018686}, which zoom into users' detailed feature-level opinions to explain the recommendations. But these solutions simply map feature-level comments into numeric ratings, a and thus ignore the detailed reason that the user likes/dislikes the feature of a particular item. For example, in Figure \ref{fig:compositions} the user endorses the mobile phone's screen because of its large size and high definition; but the phrase-level sentiment analysis only discloses the user is in favor of this phone's screen feature. 
It is impossible for such type of algorithms to explain how exactly the highlighted features of their recommendations match the user's specific preference.


We focus on explaining factorization-based recommendation algorithms \cite{Baltrunas:2011:MFT:2043932.2043988, doi:10.1137/07070111X}
by taking a holistic view of item recommendation and sentiment analysis. We develop a joint tensor factorization solution to integrate two complementary tasks of \textit{user preference modeling for recommendation} and \textit{opinionated content modeling for explanation}, i.e., a multi-task learning approach \cite{Evgeniou:2004:RML:1014052.1014067,Argyriou:2008:CMF:1455903.1455908,DBLP:journals/corr/abs-1205-2631}. 
The task of item recommendation is modeled by a three-way tensor over user, item and feature, to describe users' preferences on individual items' features, constructed by feature-level sentiment analysis in opinionated review content. The companion task of opinionated content analysis is modeled by another two three-way tensors, one over user, feature, opinionated phrase, and another over item, feature, opinionated phrase, both of which are constructed from user-generated review content. Among these two tensors, the first one specifies what kind of text descriptions a user usually provides to depict a particular feature of a given item, and the second describes what kind of opinionated comments an item often receives on its features. Via a joint factorization over these three tensors, we map users, items, features and opinionated phrases to a shared latent representation space. Item recommendation can be performed by projecting items onto the space spanned by the user factors; and explanations at different levels of granularity can be generated by projecting the features and opinionated phrases onto the space jointly spanned by the user and item factors. Cursed by the high dimensionality, sparsity is a serious issue in both learning tasks; our joint factorization method alleviates this challenge by exploiting relatedness between these two companion tasks.  

Extensive experimental comparisons between our proposed solution and existing explainable recommendation algorithms demonstrate the effectiveness of our solution in both item recommendation and explanation generation tasks in two different application scenarios, i.e., product recommendation based on Amazon reviews and restaurant recommendation based on Yelp reviews. In particular, we perform serious user studies to investigate the utility of our explainable recommendation in practice. Positive user feedback further validates the value of our proposed solution.

\section{Related work}
Studies show that explanations play an important role in helping users evaluate a recommender system \cite{swearingen2001beyond,symeonidis2008providing}. In one of the early studies of explanations in recommender systems, Herlocker et al. evaluated 21 types of explanation interfaces for a collaborative filtering based system \cite{herlocker2004evaluating} and found a histogram showing the ratings from similar users was the most persuasive. 
Bilgic and Mooney \cite{bilgic2005explaining} studied three explanation generation strategies, i.e., keyword-style, neighbor-style, and influence-style, in a content-based book recommendation system. Sharma and Cosley \cite{sharma2013social} conducted users studies to investigate the effect of social explanations, e.g., ``X, Y and 2 other friends like this.'' 
However, these studies focus mostly on content-based collaborative filtering algorithms, whose recommendation mechanism is easy to interpret but recommendation quality is known to be inferior to modern latent factor models. 

Latent factor models, such as matrix factorization \cite{Baltrunas:2011:MFT:2043932.2043988} and tensor factorization \cite{doi:10.1137/07070111X}, have achieved great success in practical recommender systems. This type of algorithms map users and recommendation candidates to a lower dimensional latent space, which encodes affinities between different entities. 
Despite their promising recommendation quality, the latent and nonlinear characteristics of this family of solutions make it frustratingly difficult to explain the generated recommendations.  



The lack of interpretability in factorization-based algorithms has attracted increasing attention in the research community.
Zhang et al. \cite{Zhang:2014:EFM:2600428.2609579,Chen2016Learning} combined techniques for phrase-level sentiment analysis with matrix factorization. 
Abdollahi and Nasraoui \cite{abdollahi2017using} introduced explainability as a constraint in factorization: the learnt latent factors for a user should be close to those learnt for the items positively rated by him/her. 
However, such type of algorithms only explain ratings, either the overall rating or feature-level ratings, while ignore the details in a user's comment. They are restricted to some generic explanations, such as ``You might be interested in [feature], on which this product performs well'' \cite{Zhang:2014:EFM:2600428.2609579}. Our work introduces a companion learning task of opinionated content modeling, in parallel with the task of factorization based recommendation. We explicitly model how a user describes an item's features with latent factors, so that we can explain why he/she should pay attention to a particular feature of a recommended item, e.g., ``We recommend this \textbf{phone} to you because of its \textit{high-resolution} \textbf{screen}.'' 

There are also solutions considering the latent factor models from a probabilistic view, which provides the flexibility of modeling associated opinionated text data for explanation. Wang and Blei \cite{wang2011collaborative} combine probabilistic matrix factorization with topic modeling for article recommendation. Explanations are provided by matching topics in items against the target user.
A follow-up work \cite{Diao:2014:JMA:2623330.2623758} introduces aspect-level topic modeling to capture users' finer-grained sentiment on different aspects of an item, so that aspect-level explanations become possible. Ren et al. \cite{Ren:2017:SCV:3018661.3018686} introduce social relations into topic modeling based recommendation via a concept named viewpoint, which enables social explanation. However, the probabilistic modeling of latent factors is usually limited by the feasibility of posterior inference, which restricts the choices of distributions for modeling the rating and text content. And the resolution of explanations is often confined at the topic level, which leads to generic explanation across all users. Our solution directly works with factorization-based latent factor models to capture a more flexible dependency among user, item and the associated opinionated content. Via a joint tensor factorization, latent representation of each opinionated phrase in the vocabulary is learnt for generating personalized context-aware explanations.   

\vspace{-2mm}
\section{Methodology}
In this section, we elaborate our multi-task learning solution for explainable recommendation. We exploit the opinionated review text data that users provide in addition to their overall assessments of the recommended items to enhance and explain the recommendation. Two companion learning tasks, i.e., \textit{user preference modeling for recommendation} and \textit{opinionated content modeling for explanation}, are integrated via a joint tensor factorization.

In the following discussions, we denote $m, n, p, q$ as the number of users, items, features and opinionated phrases in a dataset, and $a, b, c, d$ as the corresponding dimensions of latent factors for them in the learnt model. As a result, after factorization, users, items, features and opinion phrases can be represented by four non-negative matrices $U\in\mathbb{R}_{+}^{m\times{a}}$, $I\in\mathbb{R}_{+}^{n\times{b}}$, $F\in\mathbb{R}_{+}^{p\times{c}}$ and $O\in\mathbb{R}_{+}^{q\times{d}}$ in the latent factor space, respectively. Note that these four types of entities are associated with different degrees of complexity in practice, and therefore we do not restrict them to the same dimension of latent factors. 
To capture users' detailed feature-level opinions, we assume the existence of a domain-specific sentiment lexicon $\mathcal L$. Each entry of $\mathcal L$ takes the form of (feature, opinion, sentiment polarity), abbreviated as $(f, o, s)$, to represent the sentiment polarity $s$ inferred from an opinionated text phrase $o$ describing feature $f$. 
Specifically, we label the sentiment polarity $s$ as positive (+1) or negative (-1), but the developed solution can be seamlessly extended to multi-grade or continues rating cases. Based on this notation, our sentiment analysis is to map each user's review into a set of $(f,o,s)$ entries. We use $R^{U}_{i}$ and $R^{I}_{j}$ to denote the set of reviews associated with user $i$ and item $j$, respectively.

\subsection{User Preference Modeling for Item Recommendation}
\label{user preference modeling}
This task is to predict the relevance of a recommendation candidate to a user, such that relevant candidates can be ranked higher. Traditional solutions perform the estimation by mapping users and items to a shared latent space via low-rank factorization over an input user-item affinity matrix \cite{Koren:2009:MFT:1608565.1608614,Ma:2008:SSR:1458082.1458205}. However, because this input matrix is usually constructed by users' overall assessment of items, such as clicks or ratings, the learnt factors cannot differentiate nor explain the detailed reason that a user likes/dislikes an item. To address this limitation, we focus on feature-level preference modeling for item recommendation. We aim to not only predict a user's overall assessment of an item, but also his/her preference on each feature of this item to enhance the recommendation. 

Since different users would focus on different features of the same item, and even for the same feature of an item, different users might express distinct opinions on them, we use a three-way tensor ${X}\in{\mathbb{R}_{+}^{m\times{n}\times{p}}}$ to summarize such a high-dimensional relation. The key is to define the element $X_{ijk}$ in this tensor, which measures to what extent user $i$ appreciates item $j$'s feature $k$ reflected in his/her opinionated review set $R^{U}_{i}$. In this work, we adopt the method developed in \cite{lu2011automatic} to construct a domain-specific sentiment lexicon $\mathcal L$ for analyzing users' detailed feature-level opinions. 
As the construction of a sentiment lexicon is not a contribution of this work and limited by space, we will not discuss the details of this procedure; interested readers can refer to \cite{Zhang:2014:EFM:2600428.2609579,lu2011automatic} for more details.


Based on the constructed sentiment lexicon $\mathcal L$, a user review can be represented as a list of $(f, o, s)$ tuples. It is possible that a user mentions a particular feature multiple times in the same review but with phrases of different sentiment polarities. 
To denote the overall sentiment, we calculate the summation of all sentiment polarities that user $i$ has expressed on item $j$'s feature $k$. Suppose feature $k$ is mentioned $t_{ijk}$ times by user ${i}$ about item ${j}$ with the sentiment polarity labels $\{s_{ijk}^1, s_{ijk}^2, \ldots, s_{ijk}^{t_{ijk}}\}$, we define the resulting feature score as $\hat s_{ijk} = \sum^{t_{ijk}}_{n=1} s_{ijk}^n$. 

As we discussed in the introduction, a user's overall assessment of an item is usually a composition of multiple factors. In order to build the connection between a user's feature-level and overall assessments of an item, we introduce the overall assessment as a dummy feature to all items and append the overall rating matrix $A\in\mathbb{R}_{+}^{m\times{n}}$ to tensor $X$. This results in a new tensor $\tilde X\in{\mathbb{R}_{+}^{m\times{n}\times{(p+1)}}}$. To normalize the scale between feature score $\hat s_{ijk}$ and item overall rating $A_{ij}$ in $\tilde X$, we perform the following nonlinear mapping on its elements introduced by the feature scores,
\begin{equation}
  \tilde X_{ijk} = \left \{ \begin{array}{l}
  0, \text{if $f_{k}$ is not mentioned by $u_{i}$ about $i_{j}$}\\ 
  1 + \frac{N-1}{1+\exp(-\hat s_{ijk})}, \quad\textrm{otherwise}
  \end{array} \right.
\end{equation}
where $N$ is the highest overall rating in the target domain. 

Tensor $\tilde X$ describes the observed affinity among users, items and features in a training data set. To predict unknown affinity among these three types of entities in testing time, we factorize $\tilde X$ in a lower dimensional space to find the latent representation of these entities, and complete the missing elements in $\tilde X$ based on the learnt representations. As we do not restrict these three types of entities to the same dimension of latent factors, we require a more flexible factorization scheme. Tucker decomposition \cite{Karatzoglou:2010:MRN:1864708.1864727,doi:10.1137/07070111X} best fits for this purpose, i.e., 
\begin{align}
\label{eq_tucker_obj}
    &\min_{\hat X}~~||\hat X-\tilde X||_F \\
    \text{s.t.} ~~&\hat X = \sum^a_{r=1}\sum^b_{t=1}\sum^c_{v=1}g_{rtv}\mathbf{u}_r\otimes\mathbf{i}_t\otimes\mathbf{f}_v\nonumber \\
    & \forall r,t,v~~~~ \mathbf{u}_r\ge0, \mathbf{i}_t\ge0, \mathbf{f}_v\ge0, \text{and}~g_{rtv}\ge0 \nonumber
\end{align}
where $\mathbf{u}_r$ is the $r$-th column in the resulting user factor matrix $U$, $\mathbf{i}_t$ is the $t$-th column in the resulting item factor matrix $I$, $\mathbf{f}_v$ is the $v$-th column in the resulting feature factor matrix $\tilde F$ (with dummy overall assessment feature expansion), $||\cdot||_F$ denotes the Frobenius norm over a tensor, and $\otimes$ denotes vector outer product. As we have mapped the feature scores to the same range of overall ratings in the target domain (i.e., $[1, N]$), we impose non-negative constraint over the learnt latent factors to avoid any negative predictions.

In Tucker decomposition, a core tensor $\mathcal{G}\in \mathbb{R}_{+}^{a\times{b}\times{c}}$ is introduced to describe how and to what extent different tensor elements interact with each other. This provides us another degree of freedom in performing the joint factorization in our multi-task learning solution. We will carefully elaborate this important advantage later when we discuss the detailed learning procedure in Section \ref{Learning}. 

The resulting factor matrices $U$, $I$, and $F$ are often referred to as the principal component in the respective tensor mode. And the unknown affinity among user $i$, item $j$ and feature $k$ can therefore be predicted by,
\begin{equation}
\label{eq_vct_pred}
   \hat X_{ijk}=\sum^a_{r=1}\sum^b_{t=1}\sum^c_{v=1}g_{rtv}\mathbf{u}_{ri}\mathbf{i}_{tj}\mathbf{f}_{vk}. 
\end{equation}
The predicted user's feature-level assessment can already serve as a form of rating-based explanation 
\cite{Zhang:2014:EFM:2600428.2609579}. In the next section, we will enhance our explanation to free text based, by learning from user-provided opinionated content about the items and features.   


We should note recommendation is essentially a ranking problem, in which one needs to differentiate the relative relevance quality among a set of recommendation candidates. However, the current Tucker decomposition is performed solely by minimizing element-wise reconstruction error, i.e., in Eq~\eqref{eq_tucker_obj}, which cannot directly optimize any ranking loss. To address this limitation, we introduce the Bayesian Personalized Ranking (BPR) principle \cite{DBLP:journFals/corr/abs-1205-2618} into our factorization of $\tilde X$. Because we only have explicit user assessments at the item-level, we introduce the BPR principle in the overall rating predictions. In particular, for each user $u_i$ we construct a pairwise order set $D_i^{S}$ based on the observations about him/her in $\tilde X$:
\begin{equation*}
  D_i^{S} := \big\{(j,l)|\,\tilde x_{ij(p+1)}>\tilde x_{il(p+1)}\big\}
\end{equation*}
where $\tilde x_{ij(p+1)}>\tilde x_{il(p+1)}$ indicates in the given review set $R^{U}_{i}$: 1) the user $i$ gives a higher overall rating to item $j$ than item $l$; or 2) item $j$ is reviewed by user $i$ while item $l$ is not. Then the BPR principle can be realized by:
\begin{equation}
\label{eq_BPR}
  BPR\text{-}O_{PT} :=-\lambda_{\mathcal{B}}\sum^m_{i=1}\sum_{(j,l)\in{D_i^{S}}} \ln\sigma\big(\hat x_{ij(p+1)}-\hat x_{il(p+1)}\big)
\end{equation}
in which $\lambda_{\mathcal{B}}$ is a trade-off parameter and $\sigma(\cdot)$ is the logistic function. Intuitively, Eq~\eqref{eq_BPR} is minimized when all the pairwise ranking orders are maintained and the difference is maximized. By introducing it into the objective function of Eq~\eqref{eq_tucker_obj}, the decomposition is forced to not only reduce element-wise reconstruction error in $\tilde X$, but also to confine with the pairwise ranking order between items. 

Although we only impose ranking loss over the overall rating predictions in Eq~\eqref{eq_BPR}, it also implicitly regularizes the feature-level predictions. To better understand this benefit, we can rewrite Eq~\eqref{eq_vct_pred} into a matrix product form, 
\begin{equation}
\label{eq_mat_pred}
   \hat X_{ijk}=\mathcal{G}\times_a U_i \times_b I_j \times_c \tilde F_k 
\end{equation}
where $\mathcal{G}\times_n M$ denotes the $n$-mode product between tensor $\mathcal{G}$ and matrix $M$, i.e., multiply matrix $M$ with each mode-$n$ fiber in $\mathcal{G}$. 

For a given pair of user $i$ and item $j$, the first two $n$-mode product results in a matrix, denoted as $T_{ij}$, which presents a $(p+1)\times c$ dimensional space spanned by the latent factors for user $i$ and item $j$. The feature scores and overall ratings are predicted by projecting the feature factors, i.e., matrix $\tilde F$, onto it. To satisfy the BRP principle in Eq~\eqref{eq_BPR}, $T_{ij}$ has to be adjusted for each pair in $D_i^{S}$. As $\tilde F$ is globally shared across users and items, this introduces the pairwise ranking loss into the gradient for all features' latent factor learning; this effect becomes more evident when we introduce the learning procedures for our joint factorization later in the Section \ref{Learning}.

\subsection{Opinionated Content Modeling for Explanation}
\label{content_modeling}

If an algorithm could predict the opinionated content that the user would provide on the recommended item, it is an informative explanation to reveal why the user should pay attention to those features of the recommendation. Based on this principle, we develop a companion learning task of opinionated content modeling to generate detailed textual explanations for the recommendations. 

With the factorization scheme discussed in the last section, a straightforward solution for content modeling is to create a four-way tensor to summarize the complex relations among users, items, features, and opinion phrases. However, this four-way tensor would be extremely sparse in practice, as an ordinary user would only comment on a handful of items and we cannot expect their comments to be exhaustive. It is known that in natural language the distribution of words is highly skewed, e.g., Zipf's law \cite{li1992random}; we hypothesize that the distribution of opinion phrases that an item often receives for describing its features, and that a user often uses to describe a type of items' features are also highly skewed. In other words, the appearance of an opinion phrase towards a feature should strongly depend on the user or the item. Therefore, we approximate the four-way tensor by two three-way tensors, one summarizes the relation among user, feature and opinion phrase, and another for item, feature and opinion phrase.

This approximation is also supported by prior studies in mining opinionated text data. 
Amarouche \cite{Amarouche2015Product} specifies that opinion phrase associated with a feature is apparently dependent on the opinion holder (user) as well as the target object (item) in product opinion mining. Kim and Hovy \cite{Kim2005Identifying} focus on the importance of the opinion holder, explaining that the opinion holder's identification can be used independently to answer several opinion questions. Ronen and Moshe \cite{Feldman2007Extracting} compare products on their features/attributes by mining user-generated opinions, and report the dependence of opinions on different products features/attributes. In our experiments, we also empirically confirmed our hypothesis for approximation via a permutation test on two large review data sets. More details about this test will be discussed in Section \ref{offline_exp}.

We denote the first tensor as $Y^U\in\mathbb{R}_{+}^{m\times{p}\times{q}}$. From the review set $R^{U}_{i}$ of user $i$, we extract  all positive phrases this user has used to describe feature $k$ across all items, i.e., $\mathcal{R}^{U}_{i,k} = \big\{o|\,(f,o,s)\in{R^{U}_i}, f=k, s=+1\big\}$. We only include positive phrases, as we need to explain why a user should appreciate the feature of a recommended item, rather than avoid it; otherwise we should not recommend this item or feature at all. But our algorithm can be easily extended to the scenario where one needs to provide warning messages (e.g., include the negative phrases in the tensor). To reflect the frequency of user $i$ uses opinion phrase $o$ to describe feature $k$, and to facilitate the joint factorization later, we construct $Y^U$ as,  
\begin{equation}
  Y_{ikw}^{U} = \left \{ \begin{array}{l}
  0, \quad\textrm{if\ $w$\ is not\ in\ $\mathcal{R}^{U}_{i,k}$}\\ 
  1 + (N-1)\big(\frac{2}{1+\exp(-\Gamma)}-1\big),
 \quad\textrm{otherwise}
  \end{array} \right.
\end{equation}
where $\Gamma$ is the frequency of phrase $w$ in $\mathcal{R}^{U}_{i,k}$.

We construct the second tensor $Y^I\in\mathbb{R}_{+}^{n\times{p}\times{q}}$ in a similar way. For item $j$, we first obtain a collection of positive phrases about its feature $k$ from $R^{I}_{j}$, i.e., $\mathcal{R}^{I}_{j,k} =\big\{o|(f,o,s)\in{R^{I}_{j}},f=k,s=+1\big\}$, and then construct $Y^I$ as:
\begin{equation}
  Y_{jkw}^{I} = \left \{ \begin{array}{l}
  0, \quad\textrm{if\ $w$\ is not\ in\ $\mathcal{R}^{I}_{j,k}$}\\ 
  1 + (N-1)\big(\frac{2}{1+\exp(-\Omega)}-1\big), 
  \quad\textrm{otherwise}
  \end{array} \right.
\end{equation}
where $\Omega$ is the frequency of phrase $w$ in $\mathcal{R}^{I}_{j,k}$.

\begin{figure}[t]
\includegraphics[height=1.6in, width=2.5in]{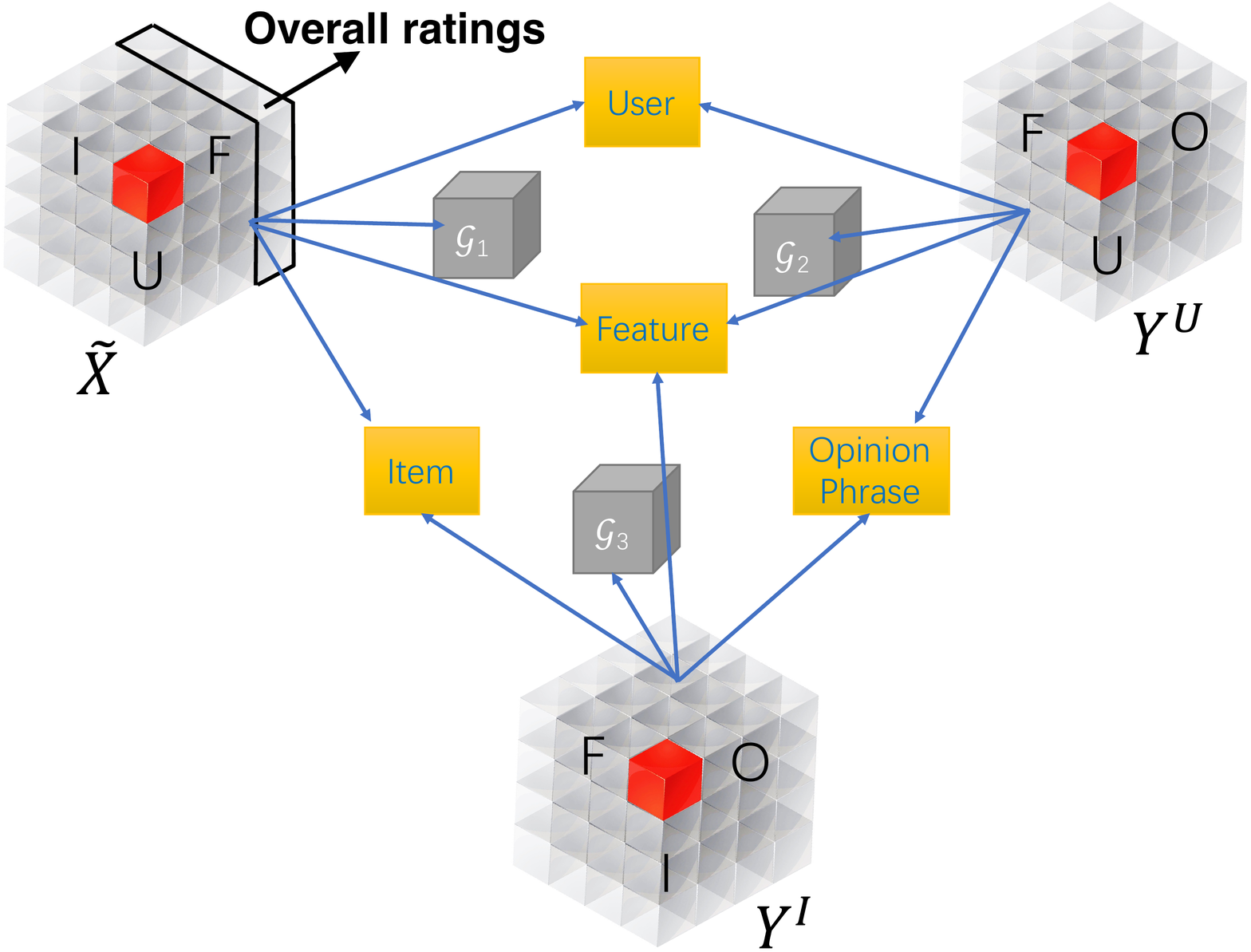}
\caption{Joint tensor decomposition scheme. Task relatedness is captured by sharing latent factors across the tensors;  in-task variance is captured by corresponding core tensors.}
\label{fig:TTTD}
\vspace{-4mm}
\end{figure}

The construction of tensor $\tilde X$, $Y^U$ and $Y^I$ impose strong dependency between the two learning tasks of item recommendation and opinionated explanation, as every two tensors share the same two types of entities (as shown in Figure \ref{fig:TTTD}). Instead of isolating the factorization of these three tensors, we propose a joint factorization scheme, which will be discussed in detail in the next section. 
Once the latent factors are learnt, we can predict user $i$'s opinionated comments on feature $k$ by the reconstructed vector $\hat{Y}^{U}_{i,k}$, which can be calculated in the same way as in Eq~\eqref{eq_vct_pred} with the corresponding latent factors. Similarly, the opinionated comments that item $j$ will receive on its feature $k$ can be predicted by the reconstructed vector $\hat{Y}^{I}_{j,k}$. As a result, to predict the opinionated comments that user $i$ will provide on item $j$'s feature $k$, we take an element-wise product between these two vectors to construct an opinion phrase scoring vector $\hat{Y}^{U,I}_{i,j,k}$, in which each element is computed as,
\begin{equation}
\label{eq_phrase_score}
  \hat{Y}^{U,I}_{i,j,k,w} = \hat{Y}^{U}_{i,k,w}\times{\hat{Y}^{I}_{j,k,w}}
\end{equation}

This estimation reflects our approximation of the original four-way tensor with two three-way tensors. Because the tensor $Y^U$ and $Y^I$ record the frequency of an opinion phrase used in describing the features by the user and about the item, Eq~\eqref{eq_phrase_score} prefers to choose those that are popularly used to describe this feature of the item in general, and also by this target user to describe this feature in similar items.  


\subsection{Multi-task Learning via a Joint Tensor Factorization}
\label{Learning}

Both of our proposed learning tasks are modeled as a tensor factorization problem, and they are coupled with the shared latent factors. Ideally, the predicted users' assessment about the recommendation candidates from the first task should be supported by the predicted users' opinionated comments about the recommendations from the second task. To leverage the dependency between these two tasks, we develop a joint factorization scheme.

In Tucker decomposition, a three-way input tensor will be decomposed into a core tensor and three principle component matrices. The core tensor captures multivariate interactions among the latent factors; and the principle component matrices can be viewed as basis of the resulting latent space. Based on this property, we decide to share the principle component matrices across the three tensors of $\tilde X$, $Y^U$ and $Y^I$ to learn the latent representations of user, item, feature and opinion phrases across the two learning tasks, and keep independent core tensors for these tensors to capture the tasks' intrinsic variance and scaling of the shared latent factors. As a result, we devise the following joint optimization formulation, 
\begin{align}
\label{objective}
\underset{\hat X, \hat Y^U, \hat Y^I}{\min}&||\hat X - \tilde X||_F + ||\hat Y^U - Y^U||_F + ||\hat Y^I - Y^I||_F \\
&-\lambda_{\mathcal{B}}\sum^m_{i=1}\sum_{(j,l)\in{D_i^{S}}} \ln\sigma\big(\hat x_{ij(p+1)}-\hat x_{il(p+1)}\big) \nonumber\\
&+\lambda_{\mathcal{F}}\big(||U||^2+||I||^2+||F||^2+||O||^2\big)\nonumber \\ &+\lambda_{\mathcal{G}}\big(||\mathcal{G}_1||^2+||\mathcal{G}_2||^2+||\mathcal{G}_3||^2\big) \nonumber\\
\text{s.t.}~~~~& \hat X=\mathcal{G}_1\times_a U \times_b I \times_c \tilde F, \nonumber\\
&\hat Y^U=\mathcal{G}_2\times_a U \times_c F \times_d O, \nonumber\\
&\hat Y^I=\mathcal{G}_3\times_b I \times_c F \times_d O, \nonumber\\
&U\ge0, I\ge0, F\ge0, O\ge0, \mathcal{G}_1\ge0, \mathcal{G}_2\ge0, \mathcal{G}_3\ge0. \nonumber
\end{align}
where we introduce L2 regularizations over the learnt latent factor matrices and core tensors to avoid overfitting. This joint factorization ensembles the two companion learning tasks for recommendation and explanation, i.e., multi-task learning; and therefore, we name our solution as Multi-Task Explainable Recommendation, or MTER in short.

The above optimization problem can be effectively solved by stochastic gradient descent (SGD), with projected gradients for non-negative constraints. However, because the number of observations in each tensor and in the pairwise ranking constraint set varies significantly, vanilla SGD procedure suffers from local optimum. To improve the convergence, we randomly select small batches of samples from each tensor and pairwise constraint set per iteration to calculate a averaged gradient, i.e., a mini-batch SGD. And to avoid manually specifying a step size, we employ adaptive gradient descent \cite{Duchi:2011:ASM:1953048.2021068}, which  dynamically incorporates the updating trace in earlier iterations to perform more informative and faster gradient-based learning.  The parameter estimation procedure of our model is off-line, and large-scale learning tasks could be solved within reasonable periods. 

Interactions between the two learning tasks become more evident in MTER when we look into the detailed gradients for model update. Denote the objective function in Eq~\eqref{objective} as $L$, and we list the gradient of $F_k$ as an example to illustrate how the elements in three tensors $\tilde{X}$, $Y^U$ and $Y^I$ contribute to it:

\begin{align}
\label{gradient}
    \frac{\partial{L}}{\partial_{F_k}} =& \frac{\partial{L}}{\partial_{\hat X_{ijk}}} \mathcal{G}_1 \times_a U_i \times_b I_j + \frac{\partial{L}}{\partial_{\hat Y^U_{ikl}}} \mathcal{G}_2 \times_a U_i \times_d O_w \\
    &+ \frac{\partial{L}}{\partial_{\hat Y^I_{jkl}}} \mathcal{G}_3 \times_b I_j \times_d O_w \nonumber
\end{align}
In Eq~\eqref{gradient}, as $F_k$ is shared across the decomposition of all three tensors, it bridges the other three components $U_i$, $I_j$ and $O_w$ in these two tasks. Similarly, the gradient of $U_i$, $I_j$ and $O_w$ also involves all the rest factors. Furthermore, the BPR constraint introduced on overall rating prediction indirectly affects the learning of $U_i$, $I_j$ and $O_w$, via gradient sharing. This also helps MTER conquer data sparsity issue when we have a large number of users, items, features and opinionated phrases to model.

\section{Experimentation}
\label{offline_exp}

In this section, we quantitatively evaluate our solution MTER in the tasks of item recommendation and opinionated content modeling, on two popular benchmark datasets collected from Amazon\footnote{http://jmcauley.ucsd.edu/data/amazon/} \cite{DBLP:journals/corr/HeM16,DBLP:journals/corr/McAuleyTSH15} and Yelp Dataset Challenge\footnote{https://www.yelp.com/dataset}. We perform extensive comparisons against several state-of-the-art recommendation and explainable recommendation algorithms. Improved quality in both recommendation and opinionated content prediction confirms the comprehensiveness and effectiveness of our solution. 

\subsection{Experiment Setup}

\noindent\textbf{$\bullet$ Preprocessing.} To verify our model's effectiveness in different application domains, we choose \textit{restaurant} businesses from Yelp dataset and \textit{cellphones and accessories} category from Amazon dataset. These two datasets are very sparse: 73\% users and 47\% products only have one review in Amazon dataset, and 54\% users only have one review in Yelp dataset. However, in the constructed sentiment lexicons, 401 features are extracted from Amazon dataset, and 1065 are extracted from Yelp dataset. It is very difficult to estimate the affinity between users and those hundreds of features from only a handful of reviews. To refine the raw datasets, we first analyze the coverage of different features in these two datasets. As shown in Figure \ref{fig:coverage}, only a small number of features (around 15\%) that are frequently covered in 90\% reviews, while most of features occur rarely in the whole datasets (i.e., Zipf's law). As a result, we decide to perform recursive filtering to alleviate the sparsity issue. First, we select features whose support is above a threshold; then, in turn, we filter out reviews that mentions such features below another threshold, and items and users that are associated with too few reviews. By fine tuning these different thresholds, we obtain two refined datasets with decent amount of users and items, whose basic statistics are reported in Table \ref{tab:informations}. 


\begin{table}
\centering
  \caption{Basic statistics of evaluation datasets.}
  \vspace{-3mm}
  \label{tab:informations}
  \begin{tabular}{| c | c  c  c  c c |}
    \hline
    Dataset & \#users &\#items &\#features &\#opinions &\#reviews \\
    \hline
    Amazon & 6,285 &12,626 &95 &591 &55,388 \\ 
    \hline
    Yelp & 10,719 &10,410 &104 &1,019 &285,346\\
    \hline
  \end{tabular}
  \vspace{-3mm}
\end{table}

\begin{table*}[]
\setlength{\abovecaptionskip}{0.cm}
\setlength{\belowcaptionskip}{0.cm}
\centering
\caption{Performance of personalized item recommendation.}
\label{tab:NDCGResults}
\begin{tabular}{|c|c|c|c|c|c|c|c|c|c|}
\hline
\multirow{2}{*}{\begin{tabular}[c]{@{}c@{}}NDCG\\ @K\end{tabular}} & \multicolumn{8}{c|}{Amazon}                                                     & \multirow{2}{*}{\begin{tabular}[c]{@{}c@{}}Improvement\\ best vs. second best\end{tabular}} \\ \cline{2-9}
                                                                   & MP     & NMF    & BPRMF  & JMARS  & EFM    & MTER-SA & MTER-S & MTER          &                                                                                             \\ \hline
10                                                                 & 0.0930 & 0.0604 & 0.1182 & 0.1078 & 0.1137 & 0.1147  & 0.1305 & \textbf{0.1362} & 15.23\%*                                                                                     \\ \hline
20                                                                 & 0.1278 & 0.0829 & 0.1518 & 0.1319 & 0.1465 & 0.1500  & 0.1610 & \textbf{0.1681} & 10.74\%*                                                                                      \\ \hline
50                                                                 & 0.1879 & 0.1614 & 0.2070 & 0.1980 & 0.2062 & 0.2095  & 0.2195 & \textbf{0.2268} & 9.57\%*                                                                                      \\ \hline
100                                                                & 0.2469 & 0.2050 & 0.2632 & 0.2489 & 0.2597 & 0.2560  & 0.2660 & \textbf{0.2752} & 4.56\%*                                                                                      \\ \hline
\multirow{2}{*}{\begin{tabular}[c]{@{}c@{}}NDCG\\ @K\end{tabular}} & \multicolumn{8}{c|}{Yelp}                                                       & \multirow{2}{*}{\begin{tabular}[c]{@{}c@{}}Improvement\\ best vs. second best\end{tabular}} \\ \cline{2-9}
                                                                   & MP     & NMF    & BPRMF  & JMARS  & EFM    & MTER-SA & MTER-S & MTER          &                                                                                             \\ \hline
10                                                                 & 0.1031 & 0.0581 & 0.1244 & 0.1187 & 0.1056 & 0.1265  & 0.1336 & \textbf{0.1384} & 11.25\%*                                                                                     \\ \hline
20                                                                 & 0.1359 & 0.0812 & 0.1634 & 0.1560 & 0.1366 & 0.1660  & 0.1789 & \textbf{0.1812} & 10.89\%*                                                                                      \\ \hline
50                                                                 & 0.1917 & 0.1366 & 0.2213 & 0.2103 & 0.1916 & 0.2137  & 0.2259 & \textbf{0.2369} & 7.05\%*                                                                                      \\ \hline
100                                                                & 0.2494 & 0.2169 & 0.2656 & 0.2590 & 0.2514 & 0.2611  & 0.2696 & \textbf{0.2764} & 4.07\%*                                                                                      \\ \hline
\end{tabular}
\\\emph{*p}-value < 0.05
\vspace{-4mm}
\end{table*}

\noindent\textbf{$\bullet$ Baselines.} To evaluate the effectiveness of our proposed explainable recommendation solution, we include the following recommendation algorithms as baselines:

\noindent\textbf{MostPopular:} A non-personalized recommendation solution. Items are ranked by their observed frequency in the training dataset.\\
\textbf{NMF:} Nonnegative Matrix Factorization \cite{Ding:2006:ONM:1150402.1150420}, which is a widely applied latent factor model for recommendation.\\
\textbf{BPRMF:} Bayesian Personalized Ranking on Matrix Factorization \cite{DBLP:journFals/corr/abs-1205-2618}, which introduces BPR pairwise ranking constraint into factorization model learning (as shown in Eq \eqref{eq_BPR}). \\
\textbf{JMARS:} A probabilistic model that jointly models aspects, ratings, and sentiments by collaborative filtering and topic modeling \cite{Diao:2014:JMA:2623330.2623758}.\\
\textbf{EFM:} Explicit Factor Models \cite{Zhang:2014:EFM:2600428.2609579}. A joint matrix factorization model for explainable recommendation, which considers user-feature attention and item-feature quality.\\
\textbf{MTER-S(SA):} Replace Tucker decomposition with canonical decomposition \cite{doi:10.1137/07070111X} in our MTER solution. 
With canonical decomposition, each decomposed matrix is represented as a summation of a shared component and a local component. For example, user factor matrix is then represented as $U_0 + U_1$, where $U_0$ is shared across three tensors and $U_1$ is only estimated for $\tilde X$. Similar decomposition structure design can be found in \cite{DBLP:journals/corr/abs-1205-2631,Zhang:2014:EFM:2600428.2609579}. If we do not allow the local components and assume all components are shared across tensors, we can get another variant of MTER, named as MTER-SA. 

\begin{figure}[t]
\setlength{\abovecaptionskip}{0cm}
\setlength{\belowcaptionskip}{-0.cm}
\includegraphics[height=1.1in, width=1.55in]{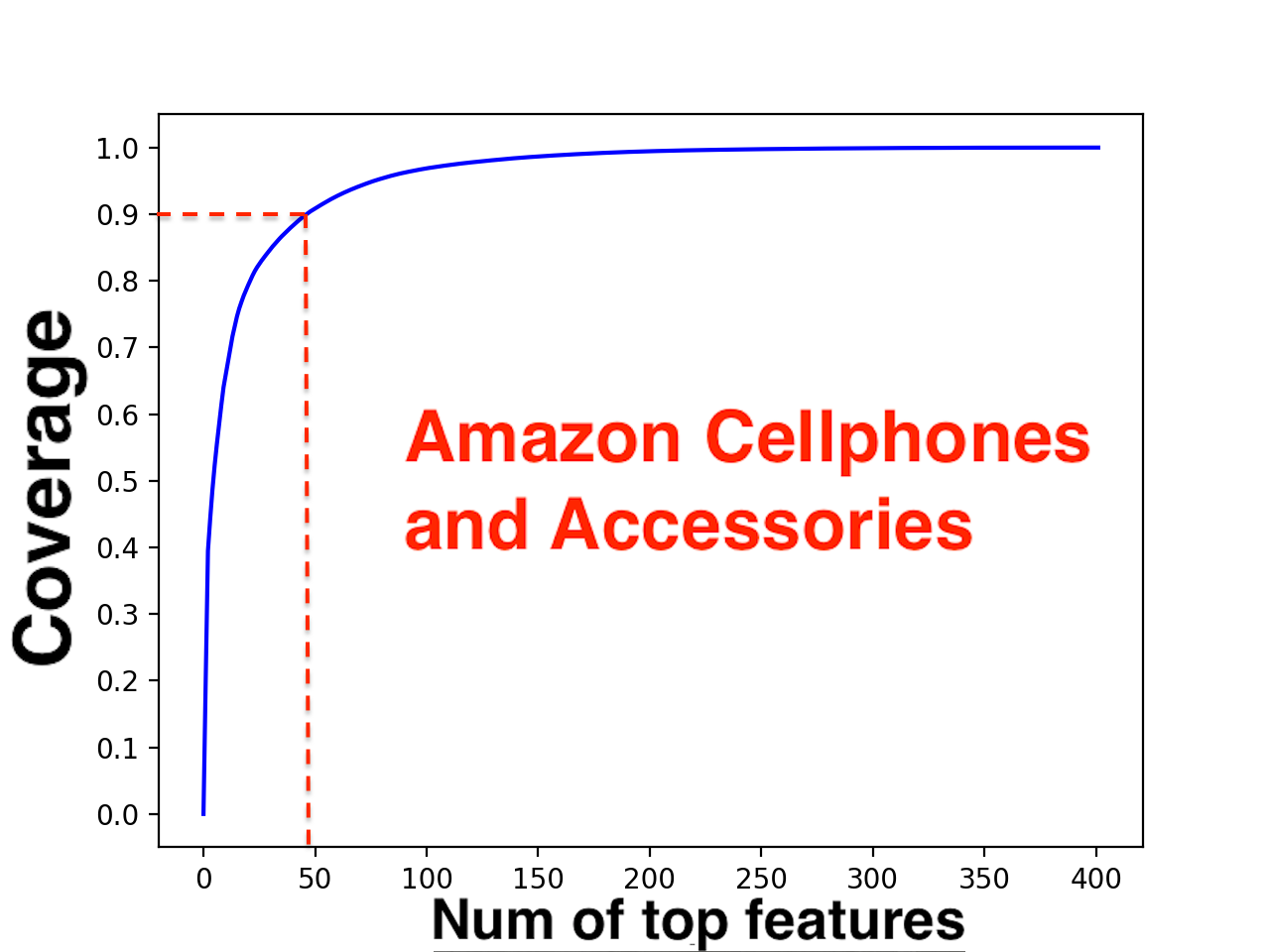}
\includegraphics[height=1.1in, width=1.55in]{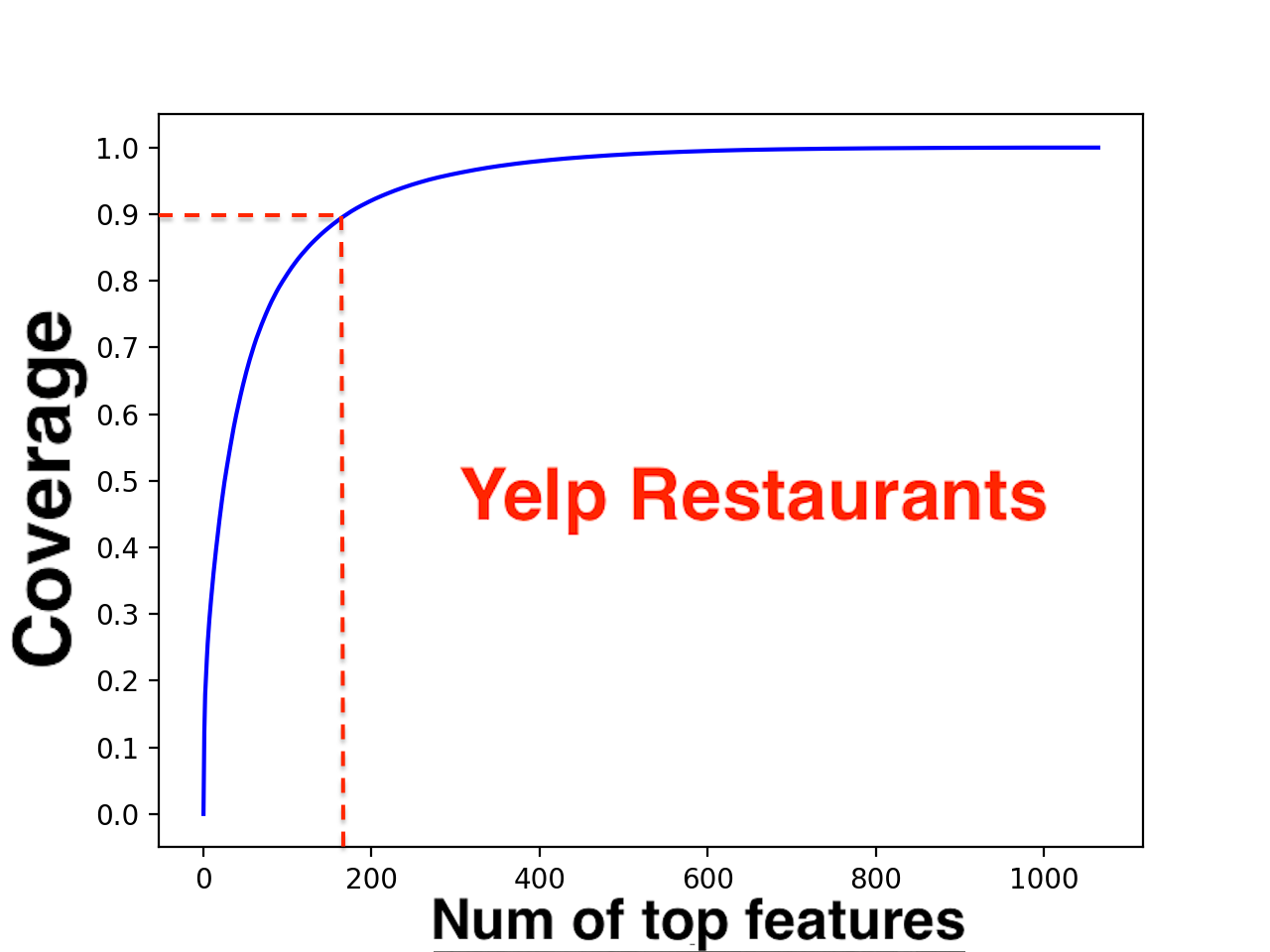}
\caption{Opinion phrase coverage test in Amazon and Yelp datasets.}
\label{fig:coverage}
\vspace{-6mm}
\end{figure} 

\noindent\textbf{$\bullet$ Evaluation Metric.}
We use Normalized Discounted Cumulative Gain (NDCG) to evaluate top-k recommendation performance of all models. 80\% of each dataset is used for training, 10\% for validation and 10\% for testing respectively. We use grid search to find the optimal hyper parameters in a candidate set for all baseline models. 

\subsection{Personalized Item Recommendation}

\noindent\textbf{$\bullet$ Performance of Recommendation.} We report the recommendation performance of each model measured by NDCG@\{10,20,50,100\} in Table \ref{tab:NDCGResults}. Paired t-test is performed between the best and second best (MTER-S(SA) excluded) performing algorithms under each metric to confirm the significance of improvement. 

\begin{figure}
\includegraphics[height=1.3in, width=1.65in]{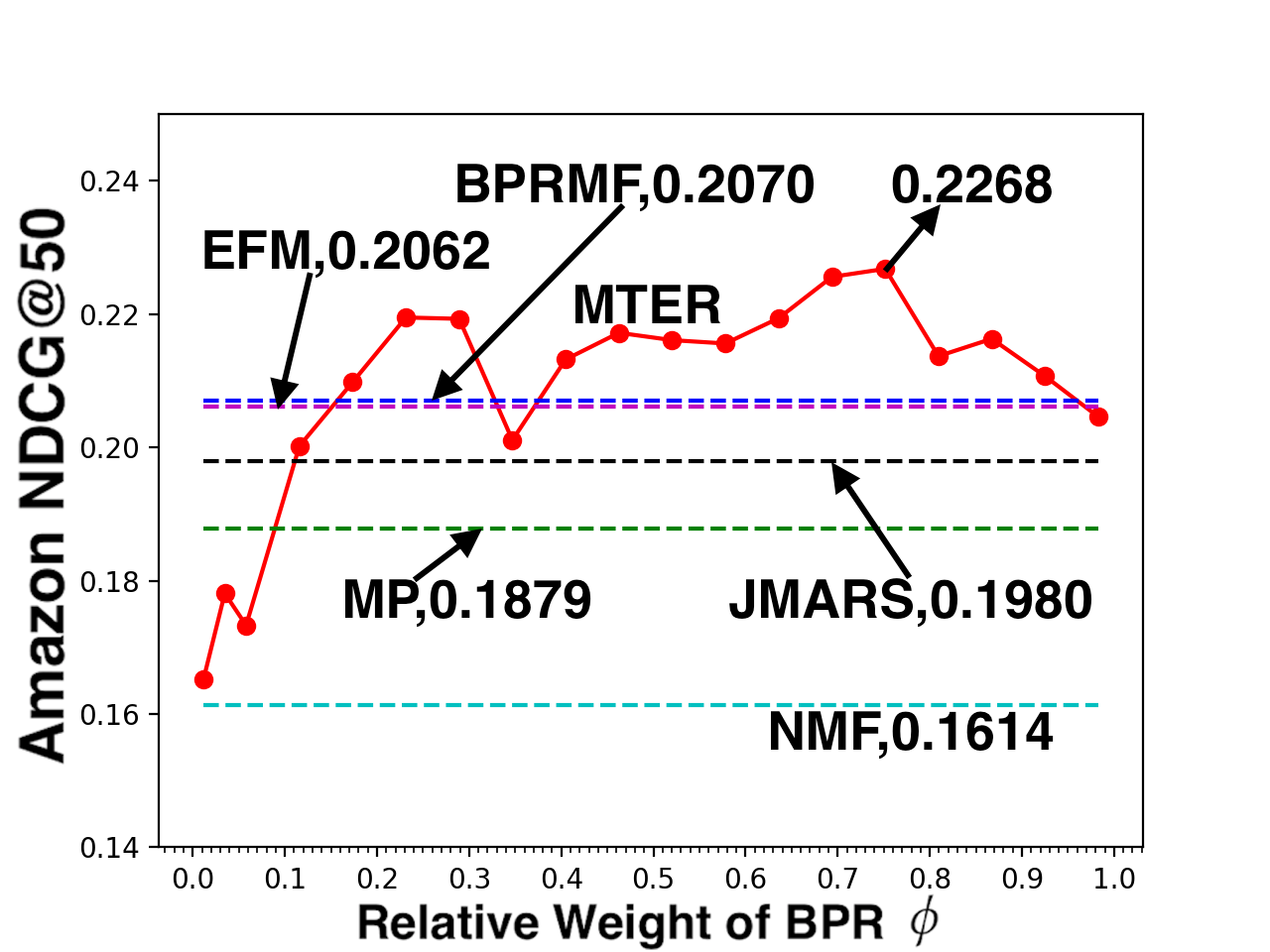}
\includegraphics[height=1.3in, width=1.65in]{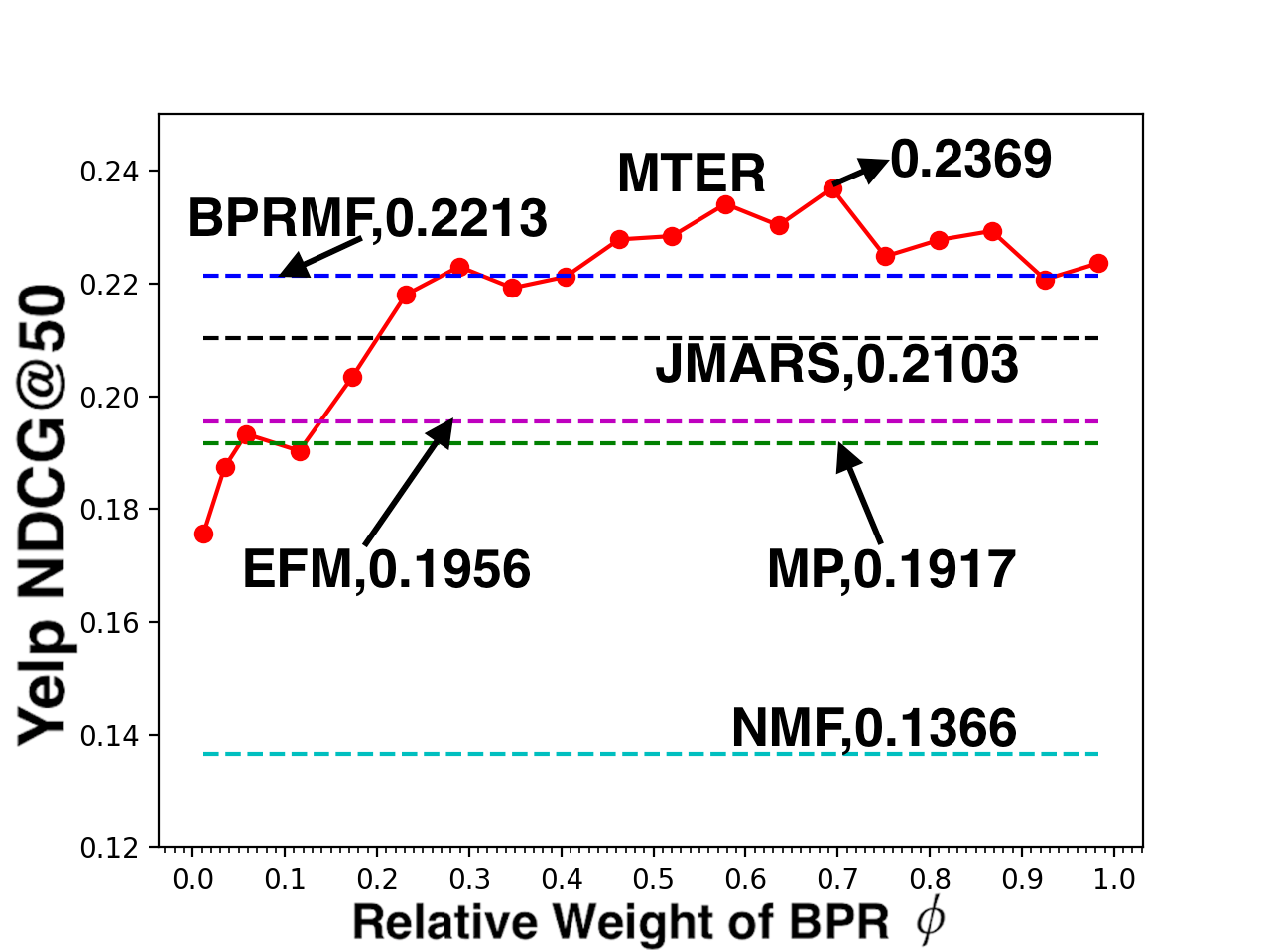}
\vspace{-3mm}
\caption{NDCG@50 $vs.$ relative weight $\phi$ of BPR on Amazon and Yelp datasets.}
\vspace{-3mm}
\label{fig:BPRWeight}
\end{figure}

Results in Table \ref{tab:NDCGResults} clearly demonstrate the advantage of MTER over the baselines. First, straightforward factorization algorithm (i.e., NMF) cannot optimize the ranking quality of the recommended items, and its performance is even worse than a simple popularity based solution, which provides generic recommendations to all users. The pairwise ranking constraints introduced by BPR greatly improve the recommendation effectiveness of BPRMF, which shares the same decomposition structure as in NMF. However, as BPRMF only models users' overall assessment on items, it cannot exploit information available in the user-provided opinionated content. Hence, its performance is generally worse than MTER and its variants. Second, comparing to JMARS and EFM, which also utilize review content for recommendation, MTER is the only model that outperforms BPRMF. JMARS models all entities in a shared topic space, which limits it resolution in modeling complex dependencies, such as users v.s., items, and users v.s., features. EFM implicitly integrates the interaction among users, items and features via three loosely coupled matrices, and it is only optimized by the reconstruction error on those three matrices. This greatly limits its recommendation quality. Third, by comparing different variants of MTER, we can recognize the advantage of Tucker decomposition in this multi-task learning setting. Because MTER-SA forces everything to be shared across three tensors, it fails to recognize task-specific variance. MTER-S enables task-specific learning, but it requires all entities to share the same dimension of latent factors. As we have observed when preprocessing the two datasets, different types of entities are associated with different number of observations, and therefore they consist of different degrees of intrinsic complexity. Forcing the latent factors to share the same structure cannot capture such intrinsic complexity, and therefore leads to sub-optimal recommendation performance. In addition, we can also observe that the best improvement from MTER is achieved at NDCG@10 (more than 15\% against the best baseline on Amazon and over 11\% on Yelp). This result is significant: it indicates a system equipped with MTER can provide satisfactory results earlier down the ranked list, which is crucial in all practical recommender systems.  

\noindent\textbf{$\bullet$ Contribution of BPR.} As the influence of BPR in our MTER training is related to both the number of pairwise constraints selected per iteration and the trade-off coefficient $\lambda_{B}$, we define a relative normalized weight of BPR to analyze its contribution in our model training: 
\begin{equation}
  \phi = \frac{\lambda_{B}\times{N_{S_{BPR}}}\times{T_{iter}}}{m\times{n}^2}
\end{equation}
where $T_{iter}$ is the the number of iterations, $N_{S_{BPR}}$ is the number of pairwise constraints sampled for BPR in each iteration, and $m\times{n}^2$ is the number of all pairwise samples from a dataset of $m$ users and $n$ items \cite{DBLP:journFals/corr/abs-1205-2618}. We fix $N_{S_{BPR}}$ and tune $\lambda_{BPR}$ for optimization. 


We evaluate NDCG@50 on an increasing weight $\phi$ for BPR, while keeping all the other hyper-parameters as constant. The result is shown in Figure \ref{fig:BPRWeight}. We can find that when $\phi$ is small, the tensor reconstruction error dominates our model learning and thus its ranking performance is worse than most baselines. But thanks to the additional information introduced in opinionated reviews, MTER is still better than NMF, which is purely estimated by the reconstruction error of the overall rating matrix. With an increased $\phi$, the pairwise ranking constraints help our model identify better latent factors that differentiate users' preferences over different items, which in turn lead to better modeling of dependency among users, items, features and opinionated phrases (as shown in Eq \eqref{gradient}). 
However, if $\phi$ goes beyond a certain threshold, the ranking quality degenerates. This is also expected: as the pairwise constraints dominate factor learning, the quality of content modeling task will be undermined, and it also increases the risk of overfitting. 

\begin{figure}
\includegraphics[height=1.2in, width=1.6in]{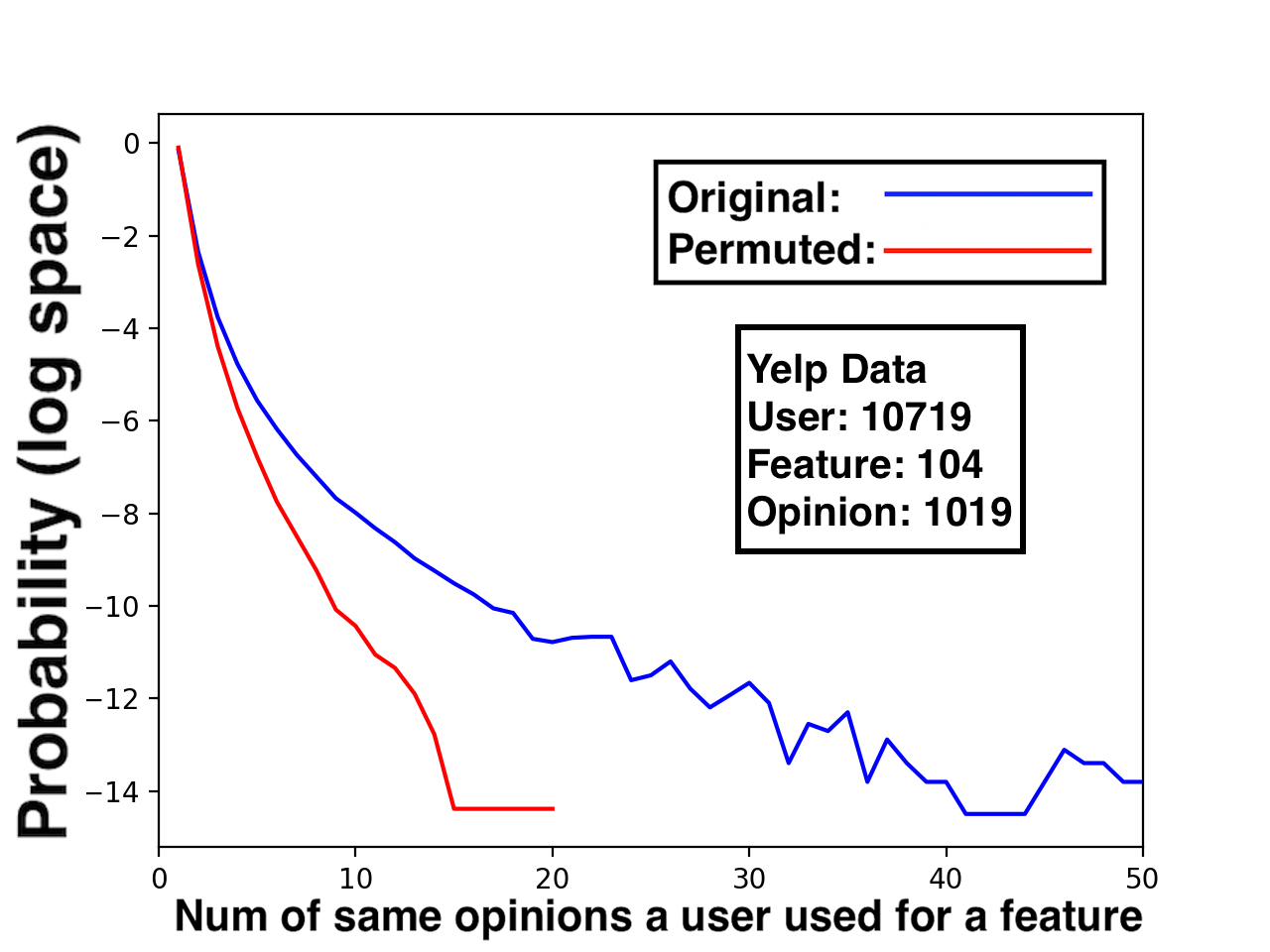}
\includegraphics[height=1.2in, width=1.6in]{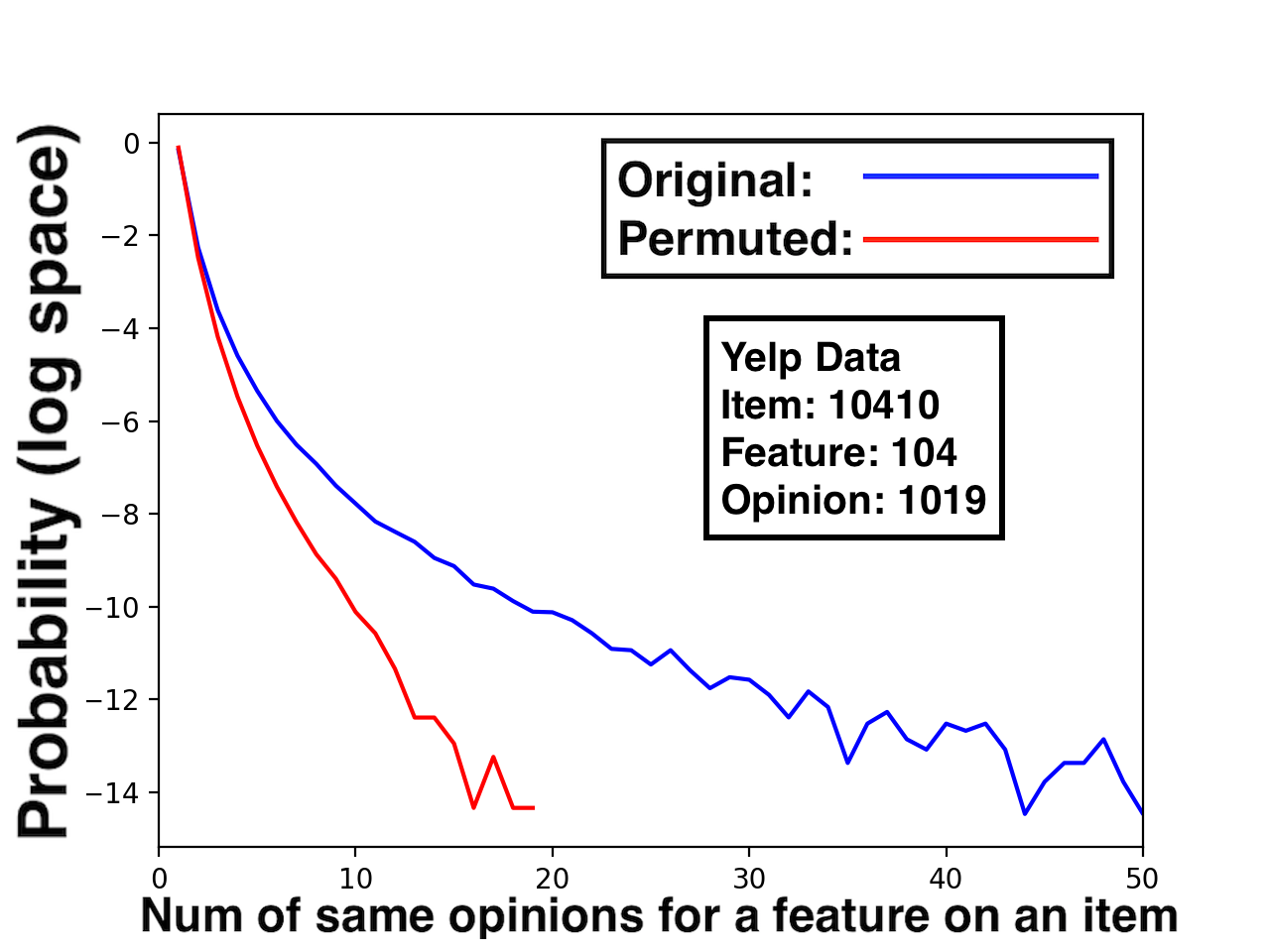}
\caption{Permutation test on the dependency between user and opinion phrase, and item and opinion phrase usages.}
\label{fig:permutation}
\vspace{-3mm}
\end{figure}

\begin{figure*}
\includegraphics[height=2.2in, width=5.3in]{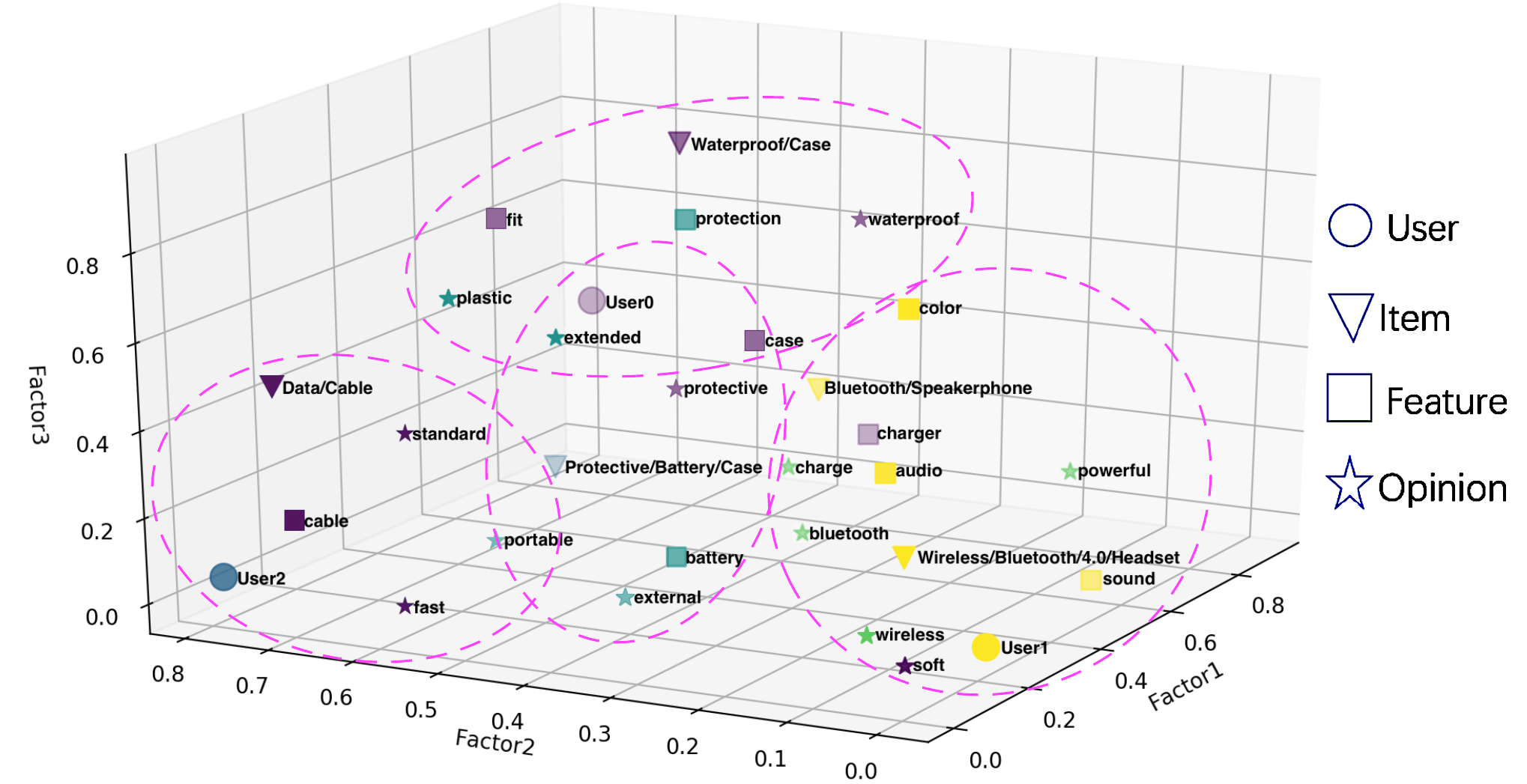}
\vspace{-2mm}
\caption{Visualization of learnt latent factors by MTER from Amazon dataset.}
\vspace{-3mm}
\label{fig:qualitative}
\end{figure*}

\subsection{Opinionated Textual Explanation}
When modeling the review content for explanation generation, we assume the distribution of opinionated descriptions about a particular feature depends on the item (summarized by $Y^I$) or the user (summarized by $Y^U$). To verify this model design, we first perform a permutation test to analyze the dependency of the appearances on opinion phrases and users and items. 


In this test, we compare the frequency that a user uses the same opinion words to describe a specific feature in the original data set versus that in the permuted data sets. The null hypothesis is that the users would randomly choose the opinion phrases to describe a particular feature, with respect to their global popularity. Therefore, the choice of opinion phrases in each user review is independent from the user. To realize this null hypothesis, in the permuted dataset, opinion words for a particular feature are randomly swapped in each user's review over a subset of words that have been used to describe this feature by all users. The permutation maintains global popularity of those opinionated phrases associated with each feature in the whole corpus. We use a similar test design to verify the dependency between item and opinion phrase as well. We perform the permutation 100 times and report the averaged results. Due to space limit, we only report the test results from Yelp dataset in Figure \ref{fig:permutation}, while similar results are obtained from Amazon dataset as well. We clearly observe the probability of a user repeatedly using the same opinion phrase to describe a particular feature is obviously higher in the original dataset than that in the permuted datasets. This reflects users' discriminatory preference of using opinion words to describe a specific feature. It is also the case in the distribution of opinion phrase an item receives to describe its specific features. This permutation test verifies our assumption and supports our tensor design in modeling the opinionated content. 


We study the effectiveness of our opinionated content modeling task by evaluating if our model can predict the actual review content a user would provide on a testing item. In particular, we focus on whether our model can recognize: 1) the features that the user would pay attention to in a given item, and 2) the detailed opinion phrases the user would use to describe this particular feature. These are two very challenging prediction tasks: even after preprocessing, the datasets are still sparse, and very few repeated observations are available for a particular combination of user, item, feature and opinion phrase (as indicated in Figure \ref{fig:permutation}). The model has to leverage observations across different users, items and features, to infer their dependency. In our experiment, we use the learnt latent factors to score all possible features associated with a given item in each user, and look into the user's review to verify if the top ranked features are indeed mentioned. Similarly, we also evaluate the ranking of all possible opinion phrases associated with a specific feature to test if our model can put the users' choice on top. As EFM is the only baseline that predicts feature-level opinions, we include it as our baseline for comparison. However, because EFM cannot predict detailed opinion phrases, we also use a simple random strategy based on the feature popularity and opinion phase popularity in target user and item as our baseline.  


We report the results in Table \ref{content_pred}. MTER achieves promising performance in ranking the features that a user will mention; this proves it identifies users' true feature-level preference, which is important for both recommendation and explanation. In the opinion phrase prediction, although it is a very difficult task, MTER is still able to predict the detailed reasons that a user might endorse the item. This indirectly confirms MTER's effectiveness in explaining the recommendation results, which will be directly evaluated in our user study later. 


\begin{table}[]
\centering
\caption{Performance of opinionated content prediction.}
\vspace{-3mm}
\label{content_pred}
\begin{tabular}{|c|c|c|c|c|}
\hline
& & Random & EFM    & MTER   \\ \hline
Feature Pred. & Amazon                     & 0.4843 & 0.5094 & \textbf{0.5176}* \\ 
 NDCG@20 & Yelp & 0.3726 & 0.3929 & \textbf{0.4089}* \\ \hline
Opinion Phrase Pred. & Amazon  & 0.0314 & -   & \textbf{0.0399}* \\ 
NDCG@50 & Yelp & 0.0209 & -   & \textbf{0.0370}* \\ \hline
\end{tabular}
\\\emph{*p}-value < 0.05
\vspace{-4mm}
\end{table}

\subsection{Qualitative Analysis of Learnt Factors}

It is necessary to examine the learnt factors and understand how the complex dependency is captured. In MTER, the learnt factors can be visualized by computing the distances between those of different entities, as shown in Figure \ref{fig:qualitative}.
As a case-study, we select three users (marked by circles), five items (triangles) 
and several related features and opinions (squares and stars) from Amazon dataset for illustration. We normalize the factors and cluster the entities into 5 clusters by $k$-means. We project them to a three dimensional space by selecting the directions with the largest variance. 

We manually examine the purchasing history of these three users: $User0$ has 27 records and 10 of them are about \textit{cases or protectors}, 7 about \textit{bluetooth headsets/speakers}, 5 about \textit{batteries/chargers}, 3 about \textit{cables} and 2 about others; $User1$ has 22 records where 9 are about \textit{bluetooth headsets/speakers}, 5 about \textit{cases}, 4 about \textit{batteries} and 4 about others; $User2$ has 17 records where 8 are about \textit{cables}, 5 about \textit{cases} and 4 about others. In the learnt latent space, we can find that the most related entities enjoy shorter distances (e.g., ``Wireless/Bluetooth/4.0/Headset'' to $User1$ and ``protection'', ``waterproof'' to ``Waterproof/Case''). Moreover, this relatedness cannot be simply inferred by frequency: ``Protective/Battery/Case'' is very close to $User0$, because his/her focus on \textit{cases} and \textit{batteries} and associated features/opinionated phrases in the review history suggests his/her potential preference on a battery case. The relations captured in this factor space depend on implicit and complex associations among entities, which is a unique strength of METR. This visualization further suggests the reason of improved recommendation performance in MTER.

\section{User Study}
We perform user study to evaluate user satisfaction of the recommendations and explanations generated by MTER. In this way, we can investigate users' overall acceptance of our proposed solution of explainable recommendation: i.e., from item recommendation to feature-based explanation, and then to opinionated explanation.

\subsection{Preparation \& Setup}
Our user study is based on the review data in Amazon and Yelp datasets. For each participant, we randomly select an existing user from the review datasets, and present this user's previous reviews to the participant to read. They are asked to infer the selected user's item attentions and opinionated content preference from these reviews. Then they will judge our provided recommendations and explanations by answering several questions from this assigned user's perspective. Admittedly, this user study is simulation based, and it might be limited by the variance in participants' understanding of selected reviews. But it is very difficult to require the participants to provide their review history, and this would also lead to possible privacy concern. We leave the user study with real-world deployment of MTER as our future work.  


We carefully design the survey questions, which focus on the following three key aspects: 1) whether the explanations improve users' overall satisfaction of recommendations; 2) whether explanations provide more information for users to make a decision; 3) which is a more effective way of explanation. Based on discussions in \cite{Tintarev:2007:SER:1547550.1547664}, we create the following five questions:
\begin{itemize}
    \item[$Q1$:] Generally, are you satisfied with this recommendation?
    \item[$Q2$:] Do you think you get some idea about recommended item?
    \item[$Q3$:] Does the explanation help you know more about the recommended item?
    \item[$Q4$:] Based on the recommended items, do you think you gain some insight of why our recommend this item to you?
    \item[$Q5$:] Do you think explanations help you better understand our system, e.g., based on what we made the recommendation?
\end{itemize}
For each question, the participants are required to choose from five rated answers: 1. Strongly negative; 2. Negative; 3. Neutral; 4. Positive; and 5. Strongly positive. We intend to use $Q1$, $Q2$ and $Q4$ to evaluate satisfaction, effectiveness, and transparency of an explainable recommender algorithm, and use $Q3$ and $Q5$ to judge if our opinionated textual explanation is more effective in this problem. Based on different baselines' recommendation performance reported in Table \ref{fig:coverage}, we choose BPRMF and EFM as baselines for comparison. As BPRMF does not provide explanations, we do not ask question $Q3$ and $Q5$ in the recommendations generated by it.




We recruit our study participants through Amazon Mechanical Turk. To perform the evaluation in a more diverse population of users, we only require the participants to come from an English-speaking country, older than 18 years, and have online shopping experience. Some sanity check questions are embedded to filter careless participants, for example providing empty recommendations and explanations for them to rate.   

As MTER can directly predict opinionated phrases for explaining the recommendations, we create some very simple templates to synthesize the textual explanations. The following are two examples of recommendation and explanations MTER generated in Amazon and Yelp datasets :\\
\noindent{$\bullet$ \textbf{Amazon} Recommendation: \textit{Superleggera/Dual/Layer/Protection/case}\\Explanation: \textit{Its \textbf{grip} is \textbf{[firmer] [soft] [rubbery]}. Its \textbf{quality} is \textbf{[sound] [sturdy] [smooth]}. Its \textbf{cost} is \textbf{[original] [lower] [monthly]}.}
}
\noindent{$\bullet$ \textbf{Yelp} Recommendation: \textit{Smash/Kitchen\&Bar}\\Explanation: \textit{Its \textbf{decor} is \textbf{[neat] [good] [nice]}. Its \textbf{sandwich} is \textbf{[grilled] [cajun] [vegan]}. Its \textbf{sauce} is \textbf{[good] [green] [sweet]}.}
}
\begin{table}[]
\setlength{\abovecaptionskip}{0.cm}
\setlength{\belowcaptionskip}{-0.cm}
\centering
\caption{Result analysis of user study}
\label{tab:user-study}
\begin{tabular}{|c|c|c|c|c|c|c|}
\hline
\multicolumn{2}{|c|}{\textbf{Amazon Dataset}}                                                                                                         & \textit{Q1}    & \textit{Q2}    & \textit{Q3}    & \textit{Q4}    & \textit{Q5}    \\ \hline
\multirow{3}{*}{\textit{\textbf{\begin{tabular}[c]{@{}c@{}}Mean \\ Value\end{tabular}}}}    & BPR                                                     & 3.540          & 3.447          & -              & 3.333          & -              \\ \cline{2-7} 
                                                                                            & EFM                                                     & 3.367          & 3.360          & 3.173          & 3.240          & 3.227          \\ \cline{2-7} 
                                                                                            & MTER                                                    & \textbf{3.767} & \textbf{3.660} & \textbf{3.707} & \textbf{3.727} & \textbf{3.620} \\ \hline
\multirow{2}{*}{\textit{\textbf{\begin{tabular}[c]{@{}c@{}}Paired \\ t-test\end{tabular}}}} & \begin{tabular}[c]{@{}c@{}}MTER \\ vs. BPR\end{tabular} & 0.0142         & 0.0273         & -              & 0.0001         & -              \\ \cline{2-7} 
                                                                                            & \begin{tabular}[c]{@{}c@{}}MTER \\ vs. EFM\end{tabular} & 0.0001         & 0.0027         & 0              & 0              & 0.0004         \\ \hline
\multicolumn{2}{|c|}{\textbf{Yelp Dataset}}                                                                                                           & \textit{Q1}    & \textit{Q2}    & \textit{Q3}    & \textit{Q4}    & \textit{Q5}    \\ \hline
\multirow{3}{*}{\textit{\textbf{\begin{tabular}[c]{@{}c@{}}Mean \\ Value\end{tabular}}}}    & BPR                                                     & 3.400          & 3.387          & -              & 3.180          & -              \\ \cline{2-7} 
                                                                                            & EFM                                                     & \textbf{3.540} & 3.473          & 3.287          & 3.200          & 3.200          \\ \cline{2-7} 
                                                                                            & MTER                                                    & 3.500          & \textbf{3.713} & \textbf{3.540} & \textbf{3.520} & \textbf{3.360} \\ \hline
\multirow{2}{*}{\textit{\textbf{\begin{tabular}[c]{@{}c@{}}Paired \\ t-test\end{tabular}}}} & \begin{tabular}[c]{@{}c@{}}MTER \\ vs. BPR\end{tabular} & 0.1774         & 0.0015         & -              & 0.0013         & -              \\ \cline{2-7} 
                                                                                            & \begin{tabular}[c]{@{}c@{}}MTER \\ vs. EFM\end{tabular} & 0.3450         & 0.0128         & 0.0108         & 0.0015         & 0.0775         \\ \hline
\end{tabular}
\vspace{-2mm}
\end{table}

\subsection{Results \& Analysis}
We conduct A/B-test among the three models in this user study, where the participants are divided into 6 groups by three models and two datasets. After filtering invalid responses, we collected 900 questionnaires, i.e., 150 for each model on each dataset. The average scores on each question and results of paired t-test are reported in Table \ref{tab:user-study}. Except for $Q1$ on Yelp dataset where EFM is slightly better than MTER, our proposed solution apparently outperforms both baselines in all aspects of user study, which is also supported by the paired t-test. Comparing MTER with BPR and EFM on $Q2$ and $Q4$, we can find the effect of MTER's comprehensive explanations. Moreover, compared with EFM on $Q3$ and $Q5$, the benefit of providing both features and more precise opinion-level explanations to illustrate user's preference is further proved. 


We also collected some user feedback (which is optional) on recommendations and explanations generated by MTER:
\begin{itemize}
    \item[-] \textit{Very interesting! Good to have those words in brackets.}
    \item[-] \textit{Clearer with those descriptive words.}
    \item[-] \textit{Extreme in thoroughness, like a consumer reports thumbnail.}
\end{itemize}
This clearly demonstrates real users' desire in having such detailed explanations of the automatically generated recommendations. And our proposed MTER algorithm is a step towards the destination.


\section{Conclusion}\label{conclusion}
In this paper, we develop a multi-task learning solution via a joint tensor factorization for explainable recommendation, which spearheads a new direction for boosting the performance of recommendations and explanations jointly. 
Both offline experiments and user study show the comprehensiveness and effectiveness of our model. 

This is the first step towards exploiting explicit opinionated content for explainable recommendation; and it creates several lines of important future works. In our current solution, the dependency among different entities is implicitly modeled via the observations in users' rating and review history. It can be further explored and more accurately modeled by incorporating  external resources, such as social network structures among users and product taxonomy among items. And our tensor structure provides necessary flexibility to do so. More advanced text synthesis techniques, such as neural language models, can be utilized to generate more complete and natural sentences for explanations. Last but not least, it is important to deploy MTER in a real-world recommender system and evaluate its utility with real user populations. 
\section{Acknowledgement}
We thank the anonymous reviewers for their insightful comments. This paper is based upon work supported by the National Science Foundation under grant IIS-1553568 and CNS-1646501.

\end{document}